\documentclass[12pt,letterpaper]{JHEP3}
\pdfoutput=1
\usepackage{cite}
\usepackage{epsfig}
\usepackage{subfigure}

\graphicspath{{Figures/}}

\title{The Strong Multifield Slowroll Condition and Spiral Inflation}

\author{I-Sheng Yang\footnote{isheng.yang@gmail.com}\\
ISCAP and Physics Department \\
Columbia University, New York, NY, 10027 , U.S.A.}

\abstract{We point out the existing confusions about the slowroll parameters and conditions for multifield inflation.  If one requires the fields to roll down the gradient flow, we find that only articles adopting the Hubble slowroll expansion are on the right track, and a correct condition can be found in a recent book by Liddle and Lyth.  We further analyze this condition and show that the gradient flow requirement is stronger than just asking for a slowly changing, quasi-de~Sitter solution.  Therefore it is possible to have a multifield slowroll model that does not follow the gradient flow.  Consequently, it no longer requires the gradient to be small.  It even bypasses the first slowroll condition and some related no-go theorems from string theory.  We provide the ``spiral inflation'' as a generic blueprint of such inflation model and show that it relies on a monodromy locus---a common structure in string theory effective potentials.}

\begin{document}

\section{Introduction}

Multifield inflation models gathered a lot of attention recently, motivated by both top-down and bottom-up concerns.  Many people suggested that slowroll inflation can happen at an environment where more than one fields are dynamically important, based on UV considerations like supersymmetry\cite{BauGre11} and some specific string theory derived models\cite{Kal07,HerTeg07,BalBer07,BerRen09}.  The multifield dynamics provide can a rich variety of potential observational signatures like the isocurvature modes\cite{SasSte95} and non-gaussianities\cite{SeeLid05,Che10}.

It is somewhat surprising that a fundamental question, which some might consider as a prior before those developments, remains to have non-uniform answers.  Basically, ``{\it What are the conditions for multifield slowroll inflation}''?  By analogy to the single field slowroll models, one might want to take the ``slowroll condition'' as a property of a given point in the multi-dimensional field space.  Such condition should be both necessary and sufficient to support a self-consistent slowroll solution which follows the gradient flow.  This is therefore an expansion of the potential at this point, and the $n$th order will be related to the $n$th derivative of $V$.  It is often truncated at $n=2$ as we will do in this paper, which shall provide us what commonly known as the first and the second slowroll conditions.

Pioneers on multifield string inflation\cite{HerTeg07,BalBer07} have attempted similar goals.  As far as we can tell, the conditions derived or used in those papers do not agree with each other, and we can show that neither of them meets our explicit criterion of sufficient and necessary conditions.  The condition one can find in the classic paper by Sasaki and Stewart\cite{SasSte95} was also not meant to be necessary.  In Sec.\ref{sec-condition} we will start by analyzing these earlier works and eventually derive the correct condition.  Our first milestone is equivalent to an equation given in a recent book by Lyth and Liddle\cite{LidLyt09}.

It is worth noting that parallel to this method of expanding the potential, slowroll conditions can be studied by another well-known method---the Hubble slowroll expansion\cite{LidPar94}.  Instead of expanding $V$, the Hubble constant $H$ is expanded as a function of $\phi$.  The usual impression of their difference is the following: The potential expansion is useful to high energy physicists who study effective potentials of complicated UV theories.  The Hubble slowroll expansion is useful to astrophysicists, since it is more naturally related to the observables, thus also known as the ``phenomenological expansion''.

The equation in\cite{LidLyt09} is kind of a hybrid, that it contains explicitly both $H$ and $V$ as functions of time or of the fields.  In this paper we try to exploit the advantages of both methods more thoroughly.  We first follow the potential expansion to the end so that our condition only involves $V$.  Then at each order we remind ourselves the corresponding ``phenomenological meaning''.  This allows us to easily recognize the correct slowroll parameters as how they enter observables.  It then becomes obvious that in the multifield context, following the gradient flow is a stronger condition than just requiring a quasi-de~Sitter solution.

This discovery implies the possibility to have a non-standard slowroll inflation.  The Hubble constant $H$ is slowly changing, but the fields do not follow the gradient flow.  This possibility was already pointed out in\cite{BraHo03} but did not receive a lot of attention.  In Sec.\ref{sec-spiral} we will push the idea further by writing down an explicit potential for it.  Since the fields no longer follow the gradient flow, this type of models eliminate the need for a small $\nabla V$, which was thought to be the an obstacle of string inflation\cite{HerKac07,JohLar08a}.  It turns out to require a special point.  One that is surrounded by a radially attractive potential and allows multi-value in the angular direction.  The radial gradient provides the centripetal force and the angular gradient balances the Hubble friction.  This combination allows the fields to spiral rapidly yet descend slowly.  The multi-value property is met by the abundant monodromy loci in the string theory moduli space\cite{DanJoh06}, and recent calculation of the effective potential suggests that some of them are radially attractive\cite{DouShe07,ShiTor08,DouTor08,Mar09,AhlGre10}.  Therefore it is very likely that such ``spiral inflation'' could be realized around those monodromy loci\cite{AhlGre12}.

Finally, it is worth mentioning that as a hindsight, there is another important advantage of the Hubble slowroll expansion which is less appreciated.  In the potential expansion, one often uses $3H\dot{\phi}_i=-\partial_iV$ to relate the field motion and the potential. So it is awkward in the formalism to describe something that does not follow the gradient flow.  In Hubble slowroll expansion, $\dot{\phi}$ is related to $H$ by just the Einstein equations and no approximations involved.  Thus, its multifield generalization\cite{NibTen00,EasGib05} in principle provides the correct conditions already.  An alternative way to reach our conclusion is to follow that approach, then at the second order analyze the possible fields trajectories and shapes of the potential.

\section{Multifield Slowroll Conditions}
\label{sec-condition}

We start by reviewing the well known slowroll conditions for single field inflation models, and reminding ourselves their phenomenological meanings.  The input are the field equation of motion,
\begin{equation}
\ddot{\phi}+3H \dot{\phi} = -V'~,
\label{eq-eom1}
\end{equation}
and the assumption that the potential dominates over the kinetic energy in the Einstein equation,
\begin{equation}
3M_P^2 H^2 = V~.
\end{equation}

The consistency of a slowroll solution requires two slowroll conditions.  We will first assume that the second derivative in Eq.~(\ref{eq-eom1}) could be ignored, which later turns out to be the second slowroll condition.  Given this assumption, we can derive the first slowroll condition,
\begin{equation}
\epsilon\equiv-\frac{\dot{H}}{H^2} =3\frac{\dot{\phi}^2/2}{V}= \frac{M_P^2 V'^2}{2 V^2}\ll1~.
\end{equation}
Its phenomenological meaning is twofold: the expansion rate $H$ changes by a small fraction during one Hubble time $H^{-1}$, also the potential energy dominates.  They are directly related by virtue of the Einstein equation.

Now, the consistency of ignoring the second derivative term implies
\begin{equation}
\left|\frac{d}{dt}\left(\frac{-V'}{3H}\right)\right|\ll |V'|~,
\end{equation}
which leads us to
\begin{equation}
\left|\frac{\epsilon}{3}-\frac{M_P^2V''}{3V}\right|\ll1~.
\end{equation}
Thus it is natural to define
\begin{equation}
\eta\equiv \frac{M_P^2V''}{V}
\end{equation}
as the second slowroll parameter, and 
\begin{equation}
|\eta|\ll1
\end{equation}
as the second slowroll condition.  Again, it has a phenomenological meaning from
\begin{equation}
\frac{1}{\epsilon H}\frac{d\epsilon}{dt} = 4\epsilon - 2\eta~.
\end{equation}
Namely, the first slowroll parameter changes by a small fraction during one Hubble time.\footnote{Note that in the Hubble slowroll expansion the second order parameter is naturally defined as something proportional to $H''$, which happens to be just $\dot{\epsilon}/(H\epsilon)$.  In this paper we will follow the potential expansion, so we always define the second order parameter as the second derivative of the potential.}

In the case of multiple fields, the only difference is in the field equation of motion,
\begin{equation}
\ddot{\phi}_i+3H\dot{\phi}_i = -\partial_i V~.
\label{eq-eomM}
\end{equation}
Here we assume canonically normalized kinetic terms.  Nontrivial field space metric will promote those partial derivatives to covariant derivatives.  Since the slowroll conditions are about a particular point in the field space around which we can always locally canonically normalize, such technicality should not bother us. 

The first slowroll condition is about the first derivative of the potential, $\partial_i V$.  Though it is a vector now, obviously we only care about its magnitude and there is no room for confusion.
\begin{equation}
\epsilon\equiv-\frac{\dot{H}}{H^2} =3\left(\frac{\dot{\phi_i}^2/2}{V}\right)= \frac{M_P^2 (\partial_iV)^2}{2 V^2}\ll1~.
\label{eq-first}
\end{equation}
It also retains the same phenomenological implications.

The second slowroll condition is about the second derivative of $V$, which is now a matrix, $\partial_i\partial_j V$. It is not directly clear which part of this matrix really needs to be small.

\subsection{Conditions Appeared in the Literature}

Here we briefly review a few papers that explicitly used certain second order conditions to construct and study multifield slowroll models.  The first one is by Sasaki and Stewart\cite{SasSte95}.  It was required that the trace of the square of the matrix $\partial_i\partial_j V$ to be small,
\begin{equation}
M_P^2\sqrt{(\partial_i\partial_j V)(\partial_j\partial_i V)}\ll V~.
\label{eq-SScond}
\end{equation}
This means the curvature in all directions have to be small, which is of course sufficient.  Clearly they did not mean it to be necessary, and it is straight forward to demonstrate why not.  Consider a potential $V(\phi)$ where the single field slowroll conditions are satisfied at $\phi_*$.  We can promote it to a two field potential by simply adding an independent orthogonal direction,
\begin{equation}
V(\phi,\psi) = V(\phi) + \frac{m^2}{2}\psi^2~.
\end{equation}
A large $m^2$ immediately ruins the condition in Eq.~(\ref{eq-SScond}), but we know at $\phi=\phi_*$, $\psi=0$, slowroll inflation will occur just as in the single field potential.  Therefore Eq.~(\ref{eq-SScond}) is a sufficient but not necessary condition.

In a paper connecting string inflation models to astrophysics\cite{HerTeg07}, it was claimed that
\begin{equation}
\eta = {\rm min \ eigenvalue} 
\left\{\frac{M_P^2(\partial_i\partial_jV)}{V}\right\}~
\end{equation}
is the second slowroll parameter, and $|\eta|\ll1$ is the second slowroll condition.  This condition turns out to be not necessary nor sufficient.  We can understand that through the following example.
\begin{equation}
V(\phi_1,\phi_2)=V_0 - \frac{m_1^2}{2}\phi_1^2 + \frac{m_2^2}{2}\phi_2^2~.
\end{equation}
In the neighborhood where $V_0$ dominates, this condition means that
\begin{equation}
|\eta|=\frac{M_P^2m_1^2}{V_0}\ll1~.
\end{equation}
For a point along the $\phi_2$ axis in this region, we have
\begin{equation}
\epsilon = \frac{M_P^2m_2^4\phi_2^2}{2V_0^2}~.
\end{equation}
However, we know choosing a small $\phi_2$ would not mean slowroll even though both conditions are satisfied.  Because whether the first slowroll parameter changes slowly, 
\begin{equation}
\frac{1}{\epsilon H}\frac{d\epsilon}{dt} = 
4\epsilon - 2\frac{M_P^2m_2^2}{V_0}~,
\end{equation}
cares about $m_2$ instead of $m_1$.  Actually, it is more appropriate to take 
\begin{equation}
\eta = \frac{M_P^2m_2^2}{V_0}
\end{equation}
here, because this is what enters observables like the spectral index, not some minimum eigenvalue which is not along the rolling direction.

In another paper on string inflation\cite{BalBer07}, the authors argued that in each vector component of Eq.~(\ref{eq-eomM}), the second derivative needs to be negligible.
\begin{equation}
\left|\ddot{\phi}_i\right|\ll|3H\dot{\phi}_i| \ \ {\rm or} \ \  |\partial_iV|
\ \ {\rm for \ each \ } i~.
\label{eq-components}
\end{equation}
This starting point is fundamentally incorrect since the components do not have a specific physical meaning unless we specify a special frame.  For example, if we rotate to a frame that $\partial_iV$ goes along one of the axis, then in all the orthogonal directions its components are zero and $\ddot{\phi_i}$ cannot be negligible.  This means by definition Eq.~(\ref{eq-components}) can never be satisfied.

\subsection{The Strong Second Slowroll Condition}

In order to get the correct second slowroll condition, we recall that fundamentally it should act as a consistency condition for the approximation
\begin{equation}
3H\dot{\phi}_i \approx -\partial_i V~.
\label{eq-SRsoln}
\end{equation}
We should start from Eq.~(\ref{eq-components}) but instead of taking its components, treat it like what it is---a vector equation.  $\ddot{\phi}_i$ being negligible means its magnitude is negligible,
\begin{equation}
|\ddot{\phi}_i| \ll |\partial_iV|~.
\end{equation}
This directly implies that the change to the vector $\dot{\phi}_i$ within one Hubble time is negligible,
\begin{equation}
\frac{|H^{-1}\ddot{\phi}_i|}{3|\dot{\phi}_i|}\ll1~.
\end{equation}
After some algebra, we have
\begin{equation}
\frac{1}{3}\left[M_P^4\frac{(\partial_iV)(\partial_i\partial_jV)(\partial_j\partial_kV)(\partial_kV)}{V^2(\partial_iV)^2} - M_P^4\frac{(\partial_iV)(\partial_i\partial_jV)(\partial_jV)}{V^3} + \epsilon^2 \right]^{1/2}\ll1~.
\label{eq-long}
\end{equation}
It is useful to introduce the following notations:
\begin{equation}
\hat{V}_1 \equiv \frac{\partial_iV}{|\partial_iV|}
\end{equation}
is the normalized direction of the first derivative (gradient) vector of $V$;
\begin{equation}
\stackrel{\leftrightarrow}{V_2} \equiv 
\frac{M_P^2(\partial_i\partial_jV)}{V}
\end{equation}
is the unitless second derivative matrix of $V$.  We can simplify Eq.~(\ref{eq-long}) to
\begin{equation}
\frac{1}{3}\left[\hat{V}_1\cdot\stackrel{\leftrightarrow}{V_2}\cdot\stackrel{\leftrightarrow}{V_2}\cdot\hat{V}_1 - 2\epsilon~\hat{V}_1\cdot\stackrel{\leftrightarrow}{V_2}\cdot\hat{V}_1 + \epsilon^2\right]^{1/2} \ll1 ~,
\label{eq-short}
\end{equation}
which is identical to Eq.(20.9) in\cite{LidLyt09}.  On top of the first slowroll parameter, we have two other terms related to the projection of $\stackrel{\leftrightarrow}{V_2}$ and $(\stackrel{\leftrightarrow}{V_2})^2$ along $\hat{V}_1$~.

It is easier to understand these two terms by going to the eigenbasis of $\stackrel{\leftrightarrow}{V_2}$.  Since it is symmetric, it will have a diagonal form.
\begin{eqnarray}
\stackrel{\leftrightarrow}{V_2} &=& {\rm Diag}\{\lambda_i\}~, \\
(\stackrel{\leftrightarrow}{V_2})^2 &=& {\rm Diag}\{\lambda_i^2\}~.
\end{eqnarray}
Note that $\hat{V}_1=\{v_i\}$ in general will not be an eigenvector, so we should have
\begin{eqnarray}
\hat{V}_1\cdot\stackrel{\leftrightarrow}{V_2}\cdot\stackrel{\leftrightarrow}{V_2}\cdot\hat{V}_1 &=& \sum v_i^2 \lambda_i^2~, \\
\hat{V}_1\cdot\stackrel{\leftrightarrow}{V_2}\cdot\hat{V}_1
&=& \sum v_i^2 \lambda_i~,
\end{eqnarray} 
with $\sum v_i^2=1$.

One may imagine two ways to satisfy Eq.~(\ref{eq-short}).  Either the first 2 terms are individually small, or they mostly cancel each other.  However, since $\epsilon\ll1$, a cancellation already implies their smallness.  So the sufficient and necessary condition is
\begin{equation}
\xi \equiv \sqrt{\hat{V}_1\cdot\stackrel{\leftrightarrow}{V_2}\cdot\stackrel{\leftrightarrow}{V_2}\cdot\hat{V}_1} = \sqrt{\sum v_i^2 \lambda_i^2} \ll1.
\label{eq-strong}
\end{equation}
We call this the {\it Strong Second Slowroll Condition} because it implies that the second term in Eq.~(\ref{eq-short}) is small, too.
\begin{equation}
\eta \equiv \hat{V}_1\cdot\stackrel{\leftrightarrow}{V_2}\cdot\hat{V}_1
= \sum v_i^2 \lambda_i \ll1~.
\label{eq-weak}
\end{equation}
Why do we still care about Eq.~(\ref{eq-weak})?  Apparently by the choice of symbol $\eta$, we do intend to identify it as the analog of the second slowroll parameter in the single field case.  The reason being that the phenomenological consequence of a slowly changing first slowroll parameter, 
\begin{equation}
\frac{1}{\epsilon H}\frac{d\epsilon}{dt} = 4\epsilon - 
2 \hat{V}_1\cdot\stackrel{\leftrightarrow}{V_2}\cdot\hat{V}_1
=4\epsilon-2\eta \ll1~,
\label{eq-firstsecond}
\end{equation}
is controlled by $\eta$.  Therefore it is $\eta$ instead of $\xi$ that enters observables like the spectral index.

It is intriguing that Eq.~(\ref{eq-SRsoln}) requires Eq.~(\ref{eq-strong}), which is stronger than the requirement of Eq.~(\ref{eq-firstsecond}).  This means that with multiple fields, {\it demanding the fields to slowly follow the gradient flow is not the only way to get a slowly changing quasi-de~Sitter solution.}  This possibility was also noticed in \cite{BraHo03}.

\section{Spiral Inflation}
\label{sec-spiral}

In order to develop a slowroll model in which Eq.~(\ref{eq-SRsoln}) does not hold, it is not wise to think about conditions for the potential $V$.  Such goal always requires the use of Eq.~(\ref{eq-SRsoln}) to simplify many equations.  A parallel technique, Hubble slowroll expansion\cite{EasGib05}, is more appropriate.  Here we will again stop at the second order which is already very informative, so we can just simplifying Eq.~(\ref{eq-firstsecond}) with the full equation of motion, Eq.~(\ref{eq-eomM}), instead of Eq.~(\ref{eq-SRsoln}).

\begin{equation}
\frac{1}{H\epsilon}\frac{d\epsilon}{dt}\approx
 2\epsilon+2\frac{\dot{\phi}_i\ddot{\phi_i}}{H\dot{\phi}_i^2}~.
\end{equation}

Now it is obvious that instead of making $\ddot{\phi}_i$ small, we can satisfy the second slowroll condition by making it almost orthogonal to $\dot{\phi}_i$. In other words, the $\dot{\phi}_i$ vector can turn rapidly while maintaining a roughly constant magnitude.  Such situation is familiar to physicists as a stable circular orbit which is easiest to analyze in the polar coordinate.\footnote{Similar techniques have been used to study sharp turns which the field temporarily leaves the slowroll trajectory\cite{Che11}, or slow turns that still follow the gradient flow\cite{CheWan09}.  It was not explicitly pointed out that one can stay out of the gradient flow yet maintain slowroll inflation.}
\begin{equation}
\mathcal{L} = \frac{1}{2}\left(\dot{r}^2+r^2\dot{\theta}^2\right)-V(r,\theta)~.
\end{equation}
Here $r$ is a field with the unit of mass, $\theta$ is a unitless field.  The equations of motion are
\begin{eqnarray}
\ddot{r}+3H\dot{r}-r\dot{\theta}^2+\frac{\partial V}{\partial r}&=&0~, \\
r^2\ddot{\theta}+2r\dot{r}\dot{\theta}+3Hr^2\dot{\theta}
+\frac{\partial V}{\partial\theta}&=&0~.
\end{eqnarray}
A rapid turning slowroll can be realized when both of these equations are dominated by their last two terms.  At zeroth order we adopt the following approximation:
\begin{eqnarray}
r&=&const.=R~, \\
3HR^2\dot{\theta}&=&-\frac{\partial V}{\partial\theta}=-c~, \\
R\dot{\theta}^2&=&\frac{\partial V}{\partial r}=\frac{c^2}{9H^2R^3}~.
\end{eqnarray}
These can be satisfied by a simple choice of potential
\begin{equation}
V(r,\theta) = V_0 + c\theta + \frac{c^2r^\alpha}{9\alpha H^2 R^{\alpha+2}}~.
\label{eq-potential}
\end{equation}
More generally $c$ can be a function of $\theta$ and $r$, as long as it does not contribute significantly to the radial derivative and changes slowly enough with $\theta$.

The intuitive way to think about this model is the following.  A radially attractive potential maintains a stable circular orbit, while a slowly descending angular spiral balances the Hubble friction.  

The first slowroll condition can be derived from its phenomenological meaning of potential energy domination,
\begin{equation}
\epsilon = -\frac{\dot{H}}{H^2} =
3\frac{R^2\dot{\theta}^2/2}{3M_P^2H^2}\ll1~,
\label{eq-firstph}
\end{equation}
which leads us to
\begin{equation}
c=3\sqrt{2\epsilon}M_PRH^2\ll3\sqrt{2}M_PRH^2~.
\end{equation}
The change of $\theta$ per e-folding is
\begin{equation}
|\Delta\theta|= |H^{-1}\dot{\theta}| = \sqrt{2\epsilon}\frac{M_P}{R}~.
\end{equation}
If we choose Planckian radius $R\gtrsim M_P$, then $\Delta\theta\ll1$ and it is not turning rapidly enough.  This recovers the usual gradient flow inflation.  On the other hand, we can choose
\begin{equation}
R\lesssim\sqrt{2\epsilon}M_P~,
\label{eq-spiralcond}
\end{equation}
such that the field rotates a significant fraction of $2\pi$ per e-folding.  That is the spiral slowroll we are looking for.

Of course, this implies that the potential is not singled-value after a $2\pi$ rotation.  This is totally fine and actually exciting.  In string theory, one can get an effective potential for the moduli fields from Calabi-Yau compactification.  The moduli space always comes with several branch-cuts and multiple layers.  The end points of these branch-cuts are monodromy loci, some of which can have attractive potential when the strong warping correction is included\cite{DouShe07,ShiTor08,DouTor08,Mar09,AhlGre10}.

Note that in this setup, $|\nabla V|$ is dominated by the radial component that supplies the centripetal force.  Consequently, it does not have to be small and can bypass the ``first slowroll condition'' in the usual sense of a small gradient.  In fact, $|\nabla V|$ is bounded from below through the fast spiraling condition Eq.~(\ref{eq-spiralcond}),
\begin{equation}
\frac{M_p^2 |\nabla V|^2}{2V^2}\geq \frac{4}{9}\epsilon^3~.
\end{equation}  

Near a strongly warped conifold, it is possible to have an attractive potential satisfying the above properties.  However it remains unknown whether there is a good chance to sustain a long period of inflation and we will try to addressed that more general problem in \cite{AhlGre12}.

\acknowledgments 
We thank Pontus Ahlgvist, Brian Greene, Lam Hui, Shamit Kachru, David Kagan and Eugene Lim for stimulating discussions.  We especially thank Andrew Liddle and David Lyth who pointed out several useful references and provided a chapter from their book.  This work is supported in part by the US Department of Energy, grant number DE-FG02-11ER41743.

\bibliographystyle{utcaps}
\bibliography{all}

\end{document}